\documentclass[final,5p,times,twocolumn]{elsarticle}
\journal{CAD}

\usepackage{natbib}
\usepackage{amssymb}
\usepackage{caption}
\usepackage{algpseudocode}
\usepackage{amsmath}
\usepackage{amsfonts}
\usepackage{multirow} 
\usepackage{color}
\usepackage{algorithm}
\usepackage{algorithmicx}
\renewcommand{\algorithmicrequire}{\textbf{Input: }}
\renewcommand{\algorithmicensure}{\textbf{Output: }}
\begin{document}

\begin{frontmatter}



\title{Surface Reconstruction with Data-driven Exemplar Priors}

\author[lab1,lab2]{Oussama~Remil}
\author[lab1]{Qian~Xie}
\author[lab1]{Xingyu~Xie}
\author[lab3]{Kai~Xu}
\author[lab1,lab4]{Jun~Wang\corref{correspauthor}}
\cortext[correspauthor]{Corresponding author}
\ead{junwang@outlook.com}
\address[lab1]{Nanjing University of Aeronautics and Astronautics, China}
\address[lab2]{Ecole Militaire Polytechnique, Algeria}
\address[lab3]{National University of Defense Technology, China}
\address[lab4]{Huaiyin Institute of Technology, China}

\begin{abstract}
In this paper, we propose a framework to reconstruct 3D models from raw scanned points by learning the prior knowledge of a specific class of objects.
Unlike previous work that heuristically specifies particular regularities and defines parametric models,
our shape priors are learned directly from existing 3D models under a framework based on affinity propagation.
Given a database of 3D models within the same class of objects, we build a comprehensive library of 3D local shape priors.
We then formulate the problem to select as-few-as-possible priors from the library, referred to as \emph{exemplar priors}.
These priors are sufficient to represent the 3D shapes of the whole class of objects from where they are generated.
By manipulating these priors, we are able to reconstruct geometrically faithful models with the same class of objects from raw point clouds.
Our framework can be easily generalized to reconstruct various categories of 3D objects that have more geometrically or topologically complex structures.
Comprehensive experiments exhibit the power of our \emph{exemplar priors} for gracefully solving several problems in 3D shape reconstruction such as preserving sharp features, recovering fine details and so on.
\end{abstract}


\end{frontmatter}

\begin{figure*}[!t] \centering
  \includegraphics[width=\linewidth]{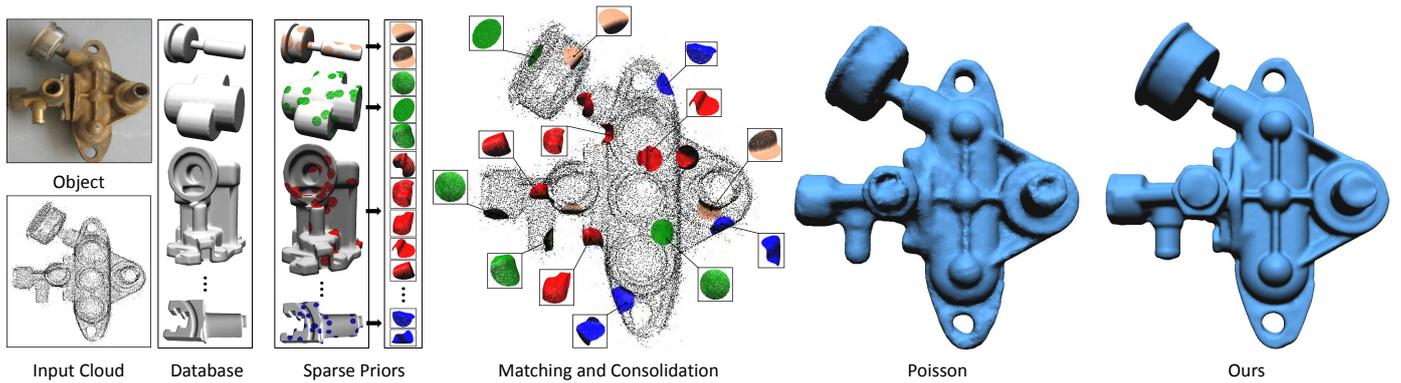}\\
  \caption{Given a noisy and sparse point cloud of structural complex mechanical part as input, our system produces the consolidated points by aligning exemplar priors learned from a mechanical shape database. With the additional information such as normals carried by our exemplar priors, our method achieves better feature preservation than direct reconstruction on the input point cloud (e.g., Poisson).}
   \label{fig:teaser}
\end{figure*}


\section{Introduction}
\label{sec:introduction}

Surface reconstruction from raw point clouds of real-world objects is of great practical importance in computer aided design and computer graphics.
Given a set of unorganized points sampled from the surface of an object, the goal is to recover the original surface.
This inverse problem is known to be inherently ill-posed if solved without any prior assumption.
Despite the recent advances in 3D scanning technology,
the point clouds acquired with scanning devices inevitably contain imperfections, such as noise and density anisotropy.
Consequently,
it makes the problem even more challenging to ensure that the reconstructed surface is geometrically and topologically faithful against
the original one with the co-existence of sharp features (such as edges and corners) and data artifacts.

Extensive study has been conducted on surface reconstruction over the past few decades~\cite{Berger2013}.
Existing methods rely on certain prior assumptions either on the sampling process, or about the nature of the surface being processed.
Those general prior assumptions, such as the smooth surface and Gaussian noise distribution,
often lead to inferior results for the raw point clouds acquired from real-world objects~\cite{Gal2007}.
Several methods take the advantage of explicitly defined priors inspired by a set of example shapes~\cite{Pauly2005,Kraevoy2005}.
Recently, a few works leverage the repetition and symmetry assumptions to extract structural priors for surface completion and reconstruction~\cite{Zheng2010,Sung2015}.
A major drawback common to these methods is that the priors are defined by heuristics and hence only suitable for particular types of objects,
e.g., planar or man-made ones.
This limitation prevents these methods from reconstructing models with more general geometry which is not captured by the pre-defined priors.

A geometric surface can be reconstructed using a combination of local patches, which can be referred to as local priors.
The number of these priors could be huge, while most of them share the same properties and they are more likely to have the same shape, specially when these priors are extracted from models within the same class of objects.
Actually, the models belonging to the same class can possess a significant number of redundant local priors sharing the same geometry, therefore these objects can be only represented by a small number of distinct local patches.
Based on this observation, we propose to use a clustering method to learn a set of local priors from a database of 3D models, rather than pre-defining heuristic reconstruction priors.
These priors are able to represent the whole class of objects from which they are extracted with much fewer parameters, and they can be used to faithfully reconstruct the raw point cloud of an object belonging to that same class, see Figure~\ref{fig:teaser}.

In our framework, we assume no prior knowledge from the shape,
and the potential priors that represent the shape of the objects are treated unknown in advance.
Our task is more general in that it extracts the priors automatically with a learning process.
Given a database of 3D models within the same class of objects, we sample point-set neighborhoods to build a comprehensive library of local shape priors from all those 3D shapes.
Our purpose is to select a representative set of priors from the library which can reconstruct any model from the same class of objects.
We formulate the problem as minimizing the reconstruction error of all database models using only the representative priors. To solve this problem, we refer to Affinity Propagation (AP) algorithm~\cite{Frey2007}, given the similarities between all prior library, it finds the set of priors that minimize the overall sum of similarities between all priors, namely \emph{exemplar priors}.
These priors are able to reconstruct the 3D shape with the same class from raw point clouds.

Since our priors are directly generated from the database models, they carry the additional geometrical information such as normals and point feature labels (i.e., regular point, edge point and corner point), and also contain abundant small features. This can be advantageous in two ways: 1) assist in the detection of sharp features from the input scan and 2) recover the fine details within the input point data even though the raw point cloud is fairly noisy and sparse. As a result, the quality of the raw point data can be significantly enhanced, which ensures the following quality surface reconstruction.

It is worth mentioning that Gal et al.~\cite{Gal2007} also proposed an example-based surface reconstruction method for scanned point sets, which utilizes a library of local shape priors built from a specifically chosen models. Their prior library is created by adding all the generated local patches from the database models. As a consequence, the number of the priors in the library could be huge, which makes the matching process quite time-consuming.
As mentioned before, the priors generated from a database within the same class could have duplicate priors that share the same geometric properties, thus the redundancy of the priors would be immoderate.
In contrast, our method learns only a set of representative priors from the library, which can be used for the reconstruction process with much less difficulty and time.

We demonstrate through experiments on the synthetic and real scan point clouds that our exemplar priors are an efficient tool to solve a variety of problems for the surface reconstruction.
To sum up, the main contributions of this paper are:
\begin{itemize}
  \item We devise a framework for surface reconstruction from existing 3D models, which requires no interaction and can easily be generalized to reconstruct various categories of 3D objects that have more complex structures from raw scanned data.
  \item We utilize a well-known clustering method (Affinity Propagation) to automatically learn the \emph{exemplar priors} from a database of 3D shapes, which can represent the 3D shape of the whole class of objects.
\end{itemize}

\section{Related Work}

\textbf{Surface reconstruction.}
Surface Reconstruction from range scanned data is a long-standing research topic which has received extensive attention from both the computer vision and computer graphics communities.
We refer the reader to the comprehensive survey on recent surface reconstruction techniques~\cite{Berger2013}.
Shape reconstruction algorithms aim to convert an input point cloud produced by aligned range scans that may have noisy and incomplete data into a high-quality surface model.
An increasing number of surface reconstruction methods merge to handle this difficult and ill-posed problem.
One class of these methods is Delaunay based surface reconstruction methods.
They interpolate the surface by taking a subset of points from an unorganized input cloud as mesh vertices,
such as the Cocone~\cite{Dey2001,Amenta2002} and the Power Crust ~\cite{Amenta2001} algorithms. 
These methods frequently produce jaggy surfaces in the presence of noise, and they require a pre-processing phase to improve their smoothness.

Another class of algorithms use surface approximation to generate a mesh from an iso-surface of a best-fit implicit function of the input point cloud ~\cite{Hoppe1992,Boissonnat2002,Kashdan2006}.
Recently, a significant effort has been placed on reconstructing surfaces with sharp features~\cite{Guennebaud2007,Daniels2007,Lipman2007,Oztireli2009}. 
This type of methods can handle data imperfections to some extent in the presence of high quality normals.
However, normal estimation and orientation are challenging problems, especially when the point data are corrupted.
In contrast, our algorithm directly uses the normals of the existing models for the input raw scans,
which usually can be accurately estimated from the triangular mesh models.
Another difficulty is to preserve sharp features of the input point data for those approaches, they can preserve the sharp features in the face of a low level of noise.
However, they have difficulties in obtaining satisfactory results in the presence of a reasonably high level of noise.

\noindent \textbf{Point cloud consolidation.}
Point cloud consolidation aims to clean up raw input data, removing a variety of data artifacts, and to provide essential geometric attributes.
A variety of algorithms have been proposed for this purpose to benefit processing tasks such as sampling and surface reconstruction.
For example, Schall et al.~\cite{Schall2007} presented a noise removal method on the static and time-varying range data.
Rosman et al.~\cite{Rosman2013} introduced a framework for point cloud denoising by patch-collaborative spectral analysis.
The Moving Least Squares (MLS) based methods can handle non-uniform data and noise,
and are ideally designed to produce smooth surfaces~\cite{Levin2003,Alexa2003}.
The variants of MLS have been developed to handle sharp features by fitting a local parametric surface at each point from the input cloud~\cite{Guennebaud2007,Oztireli2009}.

Recently, a locally optimal projection (LOP) operator was introduced by Lipman et al.~\cite{Lipman2007}.
Given an input noisy point cloud, it outputs a new point set which more faithfully adheres to the underlying shape.
On this basis, Huang et al.~\cite{Huang2009} modified and extended the LOP (weighted locally optimal projection) to deal with non-uniform distributions.
These methods often fail to recover small features and details when the input data are fairly sparse and severely noisy.
Comparatively, on the basis of the \emph{exemplar priors}, our method is capable of augmenting and consolidating the raw point data such that the data quality gets significantly enhanced. As a result, the fine details and small features can be recovered successfully.

\noindent \textbf{Non-local filtering.}
Non-local filtering methods has gained interest over the past decade, Previous works in the image processing field such as~\cite{Buades2005} attempt to tackle this problem, by introducing the notion of non-local means (NLM). The idea behind is to denoise a pixel by exploiting pixels of the whole image that may entail the same information.
This idea was also introduced for surfaces, defined as meshes or point clouds~\cite{Jones2003,Yoshizawa2006,Oztireli2009,Adams2009, Digne2012} where they extend the non-local means concept to mesh and point cloud denoising.
Schnabel et al.~\cite{Schnabel2009} presented a hole filling method
where the reconstruction is guided by a set of primitive shapes which has been detected on the input point cloud such as planes and cylinders.
Zheng et al.~\cite{Zheng2010} proposed a scan-consolidation approach to enhance and consolidate imperfect scans of urban models using local self-similarity.
Guillemot et al.~\cite{Guillemot2012} introduced a non local point set surface model that is able to exploit the surface self-similarity.

More related to our work,~\cite{Hubo2008, Digne2014} proposed a compression technique that exploits self-similarity within a point-sampled surface. In ~\cite{Hubo2008} the patches are clustered by similarity and replaced by the representative patches of each cluster, while in ~\cite{Digne2014} each patch is represented as a sparse linear combination over a dictionary.
Our work investigates the same non-local idea as~\cite{Gal2007} by using the self-similarity patches to enhance the input cloud data and reconstruct a faithful surface. Instead of using the whole library of patches, our method learns only a set of exemplar priors permitting much less difficulty and time.

\begin{figure*}[!t] \centering
  \includegraphics[width=0.95\linewidth]{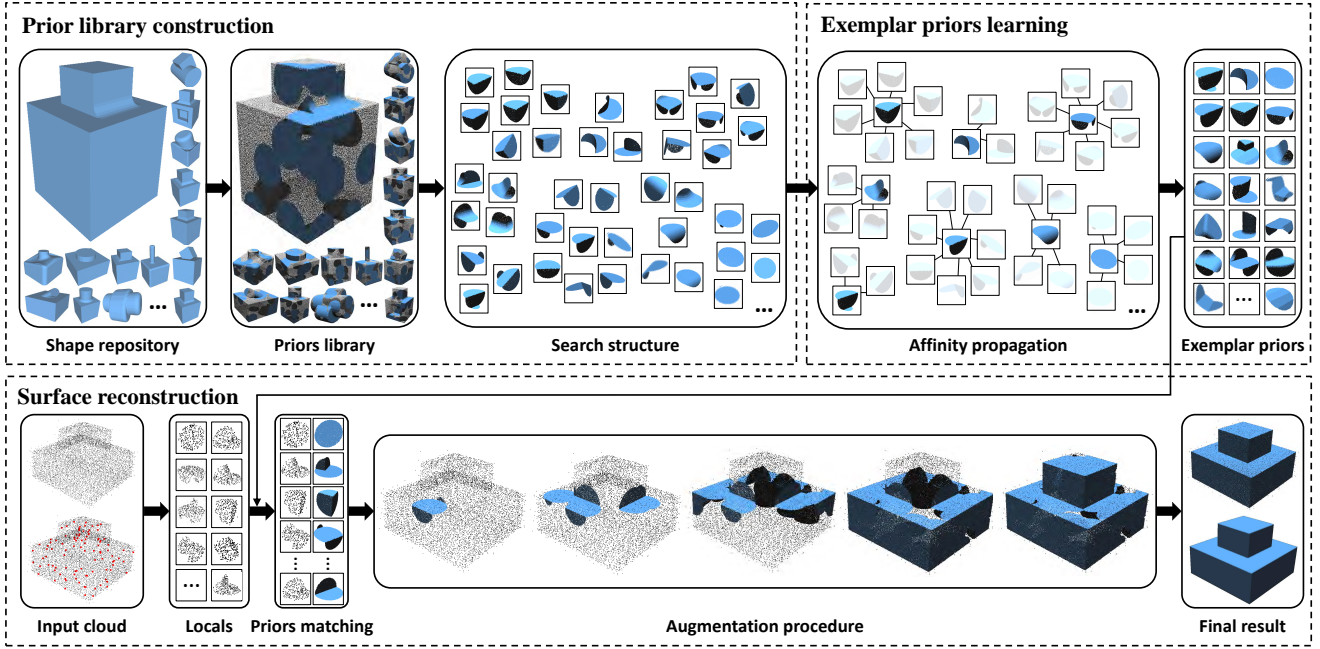}\\
  \caption{An overview of our algorithm. We extract priors from a 3D shape database within the same category (e.g., mechanical parts) to construct a prior library. The affinity propagation clustering method is then performed on the prior library to obtain the set of representative priors, called the \emph{exemplar priors}. Given an input point cloud, we construct its local neighborhoods and perform priors matching to find the similar exemplar prior to each local neighborhood.  Subsequently, we utilize the matched exemplar priors to consolidate the input point scan through an augmentation procedure, with which we can generate the faithful surface where sharp features and fine details are well recovered.}
  \label{fig:overview}
\end{figure*}
\section{Overview}

Our proposed method aims to learn a set of exemplar priors from a given 3D shape repository,
and use them to reconstruct surfaces from raw scanned point data.
The shape repository usually consists of a collection of 3D models with the same shape category.
In our work, we explore the 3D shapes from available repositories such as the PSB benchmark~\cite{Chen2009}, the Co-Seg dataset~\cite{Wang2012} and the scape database~\cite{Scape2005}.
Our algorithm comprises three major phases, as illustrated in Figure~\ref{fig:overview}.

In the first phase, we build an initial library of 3D shape priors from a set of database models.
Specifically, the 3D shapes are sampled, the seed centers of the priors are determined and the 3D shape priors are created within a bounding sphere of a predefined radius.
The geometric information such as shape descriptors, normals, labels for points belonging to sharp features, are extracted and stored for all the priors in Section~\ref{Pre-Processing}.
The priors are then inserted into a search structure based on their shape descriptors.

The initial library consists of a large amount of shape priors, which may contain numerous redundant elements.
To address this issue, we aim to extract the most representative priors from the library. We formulate the problem as minimizing the reconstruction error of all database models using only the representative priors. For that purpose, we perform affinity propagation~\cite{Frey2007}, which given the similarities between all priors, it selects some exemplar priors that minimize the overall sum of similarities between all the priors and the exemplar ones (see Section~\ref{Optimization}).
As a result, a small number of exemplar priors are obtained.

Once the exemplar priors are selected from the library, they can be utilized to augment an input raw point scan from the same class of objects where the priors are created in the first place.
Given an input cloud, we first select the seed centers for the local neighborhoods in such a way all the input cloud is covered, and then we construct the local neighborhoods within the same bounding sphere used to create the priors.
Priors matching and alignment procedures are then performed to match each local neighborhood to its nearest exemplar prior, transform the exemplar prior to the local frame, and finally register the corresponding prior onto the input scan using ICP algorithm (see Figure~\ref{fig:overview}).
Consequently, the input raw point cloud is augmented using the exemplar priors and is well consolidated.
By leveraging the moving least squares technique, we are able to generate the faithful surface to the input data, where sharp features and fine details are successfully recovered (see Section~\ref{sec:Reconstruction}).

\section{Algorithm}

In this section, we discuss every phase of the proposed algorithm in more detail. We first construct a library of 3D shape priors from a given set of database models, we compute prior descriptors, and perform Affinity Propagation to identify the subset of exemplar priors. Based on these priors only, we consolidate the raw point clouds within the same class of the input 3D shapes and reconstruct faithful surface models.

\subsection{Prior Library Creation}
\label{Pre-Processing}
\indent \textbf{Prior library.}
A prior \emph{Pr} of a 3D model $\mathcal{M}$ is defined as the set of points $p\in \mathcal{M}$ lying within a sphere of radius $R$ and a center $c$.
Given a 3D shape repository with the assumptions that the models are represented as triangle meshes and they belong to the same class, we construct a library of such local priors, and use them as input for the learning process.
The first step is to sample a set of points $P$ uniformly for each model, then we select a subset $S$ of seed centers $c_{j}$ from $P$ which satisfy the following property: the seed centers are used to create the local priors within a bounding sphere of radius $R$ in such a way the point cloud $P$ is totally covered.

The seed centers are selected in a dart-throwing fashion same as~\cite{Digne2014}. Given the sampled set of point $P$ for each model, we pick a point as a first center seed, label all its neighborhoods within a radius $R$ as covered, and traverse $P$ until a non-covered point is found and a new seed center is added to $S$, the process repeats until all points $P$ are covered. Note that the radius $R$ is chosen as a relative value to the diagonal of the bounding box of each model.
The priors are then denoted by:
\begin{equation}
\emph{Pr}(c_{j})=\{p| p \in P, c_{j}\in S , \|c_{j}-p\|\leq R\}
\label{Prior eq}
\end{equation}
where $\|c_{j}-p\|$ is the geometric distance between the center $c_{j}$ and the point $p$.

Once the priors are constructed, we store the information of each prior such as normals and points belonging to sharp features for later use, we perform weighted Principal Component Analysis (wPCA) for each prior, in such a way the principal axes coincide with the coordinate system axes and its center of mass is translated to the origin, in addition, we scale the largest principal component to 1 for all priors. Given a database of models, we are able to construct our initial prior library within the same standard coordinate system.

\textbf{Shape descriptor.}
Before we cluster similar priors into groups, we require to measure how the priors are similar to each other, in another word, we need to define a shape descriptor to present each prior.
Similar to~\cite{Gal2007}, we introduce the geometric moments as our priors descriptor thanks to their efficient and easy computation.
Since our library priors are formed by discrete points, we yield an approximation to the integral form of the moments.

Given a set of points $\textbf{X}=\{x_{i},y_{i},z_{i}\}_{i=1}^N$, their $(p,q,r)$ moment is approximated with the summation over all the points.
In order to make the moment-based representation invariant to the number of vertices of each prior, we introduce a weight $\omega_{i}$ for each point $\{x_{i},y_{i},z_{i}\}$ that is proportional to the area of the polygonal face~\cite{Bimbo2006} where that point is sampled. More formally:
\begin{equation}
M_{p,q,r}(\textbf{X})=\frac{1}{N} \sum_{i=1}^N \omega_{i} x_{i}^p y_{i}^q z_{i}^r
\label{Moment eq}
\end{equation}

We then choose a subset of moments $\{M_{p,q,r}\}$ up to an order $d$ as our prior descriptor, and obtain the vector moments $V(\emph{Pr})$:
\begin{equation}
V(\emph{Pr}) = \{ M_{0,0,1}(\emph{Pr}), M_{0,1,0}(\emph{Pr}),\ldots,M_{d,0,0}(\emph{Pr})\}.
\label{descrip eq}
\end{equation}
In our implementation, we choose a sequence of moments up to $d=6$, resulting in a $83$-dimension descriptor vector.
As a result, we are able to compute the corresponding descriptors for all priors in the library, see Algorithm~\ref{Construct priors alg}.

\begin{algorithm}
\algorithmicrequire Database models $\mathcal{M}_{i}$, $i=\{1,\ldots ,m\}$.\\
\algorithmicensure Priors library $\mathcal{L}$.
\caption{Prior library creation} \label{Construct priors alg}
\begin{algorithmic}
\Function {Prior library creation} {$\mathcal{M}_{i},R$}
   \For{$i = 1$ to $m$}
      \State $P\gets \mathcal{M}(i)$;               \Comment sample points and store normals and feature points.
      \State $S\gets$ subset$(P)$;                  \Comment find seed centers for priors.
      \For{$j = 1$ to size($S$)}
          \State $Pr(c_j)\gets$ Equation~\ref{Prior eq};           \Comment generate the prior.
          \State $Pr(c_j)\gets$ Weighted PCA $({Pr}(c_j))$;
          \State $Pr(c_j)\gets$ Canonical scaling $({Pr}(c_j))$;
          \State $V(Pr(c_j))\gets$ Equation~\ref{descrip eq};      \Comment compute the descriptor.
          \State $\mathcal{L}\gets \{Pr(c_j), V(Pr(c_j))\}$;           \Comment add the prior and its descriptor to the library.
      \EndFor
   \EndFor
   \State \Return ($\mathcal{L}$);
\EndFunction
\end{algorithmic}
\end{algorithm}

\subsection{Exemplar Priors Learning}
\label{Optimization}
Since the prior library is created from 3D models belonging to the same category of shapes, it may contain a significant number of redundant priors.
In this section, we aim to extract a relatively small number of representative priors or exemplars from the library using the affinity propagation (AP) method.

\textbf{Problem formulation.}
Given a library of priors $\mathcal{L}$, our purpose is to reconstruct all database models by using only a subset of exemplars priors from $\mathcal{L}$, denoted by $\mathcal{L}^{\star}$, which minimizes the cost function defined by the reconstruction error of all these models.
Specifically, we take advantage of the similarity measure between each pair of priors to identify the exemplar priors that minimize the standard squared error measure of similarity.
Fortunately, this problem can be solved by the affinity propagation method that recursively transmits real-valued pairwise data point similarities until a good set of exemplar data is obtained. It searches for clusters so as to minimize an objective function, i.e., the reconstruction error.

Let $\mathbf{\mathcal{M}}_{i=1}^{m}$ be the set of $m$ database models and $\mathcal{L}^{\star}$ the set of $k$ exemplar priors, i.e., $\mathcal{L}^{\star} \in \mathcal{L}$.
To reconstruct all the input models $\mathbf{\mathcal{M}}_{i=1}^{m}$, we search for each prior in $\mathcal{L}$ the most similar prior from $\mathcal{L}^{\star}$, we then align the corresponding priors from $\mathcal{L}^{\star}$ onto their similar priors from each model $\mathcal{M}_i$ to form an entire model, $\hat{\mathcal{M}_i}$.
Hence, we define the reconstruction accuracy as the geometric distance between the original model $\mathcal{M}_i$ and the reconstructed model $ \hat{\mathcal{M}_i}$ using the exemplar priors from $\mathcal{L}^{\star}$.

\begin{equation}
\mathcal{E}_i= \left\| \hat{\mathcal{M}_i} - \mathcal{M}_i \right\|
\label{Error eq}
\end{equation}
Since the geometric moments can inherently represent the shape, we replace the geometric distances between models $ \hat{\mathcal{M}_i}$ and $ \mathcal{M}_i$ with the sum of the Euclidean distances between the geometric moment vectors of their priors.

\textbf{Affinity propagation.}
In the last decade, many clustering methods have been proposed. One particularly elegant and powerful method is the affinity propagation (AP) proposed by ~\cite{Frey2007}, which can be adapted to solve our selection problem.
It is an unsupervised learning algorithm for exemplar-based clustering that finds a set of data points that best describe the input data based on the similarities between them. It associates each data point with an exemplar within the input data in a way to maximize the overall sum of similarities between data points and their exemplars.

The algorithm takes as input a collection of real-valued similarities between the priors, where the similarity $s(i,j)$ between priors $i$ and $j$ indicates how well prior $j$ is suited to be the exemplar for prior $i$, hence the similarity measure $s(i,j)$ is set to the negative Euclidean distance between the geometric moments of priors $i$ and $j$:
\begin{equation}
s(i,j) = - \left\| V(\emph{Pr}_i)- V(\emph{Pr}_j) \right\|, i\neq j
\label{eq:similarity}
\end{equation}

Suppose that there are $n$ priors in the library, each of which has the geometric moments expressed by an $83$-dimension vector.
We construct the ${n\times n}$ similarity matrix $s$ by applying equation~(\ref{eq:similarity}) for each pair of priors in the library.

Affinity propagation does not require the number of cluster to be specified. It takes as input real values for all $s(j,j)= p$ called as preferences, and these values influence the number of exemplar priors (a prior $j$ with a high preference value $s(j,j)$ is more likely to be chosen as an exemplar). In case all the priors are equally suitable as exemplars, their preferences should be set to a shared value such as the median of the input similarities, which results in a moderate number of exemplars (see ~\cite{Dueck2009} for more detail).

To decide which priors are exemplars, two kinds of message are passed between the priors in the library: (1) the responsibility $r(i,j)$ sent from prior $i$ to candidate exemplar $j$, representing how well prior $j$ is suited to be the exemplar of prior $i$ and (2) the availability $a(i,j)$ sent from candidate exemplar $j$ to prior $i$, indicating how well for prior $i$ to choose prior $j$ as its exemplar. Both messages are updated during each iteration until convergence by using the following equations:
\begin{equation}
\forall i,j: r(i,j) = s(i,j) - \max\limits_{j',j'\neq j} \{s(i,j')+a(i,j')\}
\label{eq:responsibility}
\end{equation}

$\forall i,j$: for $j=i$:
\begin{equation}
{a(i,j) = \sum\limits_{{i',i'\neq i}}\max \{ 0,r(i',j)\}}
\label{eq:availab1}
\end{equation}
$\forall i,j$: for $j\neq i$:
\begin{equation}
a(i,j) = \min \Big\{ 0,r(j,j),\sum\limits_{i',i'\notin \{ i,j\}}\max \{ 0,r(i',j)\} \Big\}
\label{eq:availab2}
\end{equation}

we assign to each prior $i$ a variable $\hat{c}_i$, a value of $\hat{c}_i=j$ for $i\neq j$ means that the prior $i$ is assigned to the cluster where $j$ is an exemplar prior; and $\hat{c}_j=j$ indicates that the prior $j$ serves as an exemplar. At any point of the algorithm, responsibilities and availabilities can be combined to make the exemplar decision and identify the cluster assignments $\hat{c}_i$ $(i=1,\ldots,n)$ defined as:
\begin{equation}
\hat{c}_i = \arg\max_{j} \{a(i,j)+r(i,j)\}
\label{eq:assignments}
\end{equation}

\textbf{Convergence.}
The algorithm can be monitored by imposing certain terminal conditions: a maximal number of iterations, the changes in the messages are below a threshold or the local decisions stay constant for some number of iterations.
We initialize the availabilities to zero and for each iteration, we (1) update the responsibilities given the availabilities; (2) update the availabilities given the responsibilities; (3) combine both of them to monitor the exemplar decision; (4) terminate the algorithm when the decisions do not change for 10 iterations, see Algorithm~\ref{AP clustering alg}.

\begin{algorithm}
\algorithmicrequire Prior library $\mathcal L$ with $n$ priors.\\
\algorithmicensure Indices of the exemplar priors  $Ind$.
\caption{Exemplar Priors Learning} \label{AP clustering alg}
\begin{algorithmic}
\Function {Learning Priors} {$\mathcal L$}
   \For{$i = 1$ to $n$}
       \For{$j = 1$ to $n$,$j\neq i$}
          \State $s(i,j)\gets$ equation~(\ref{eq:similarity});  \Comment construct similarity matrix
       \EndFor
   \EndFor
   \State $p\gets$ $median(s)$;                   \Comment the median of $s$ as a vector of preferences \For{$i = 1$ to $n$}
          \State $s(i,i)\gets p(i)$;
   \EndFor
   \State $\forall i,j: a(i,j) =0$;                 \Comment initialize availabilities.
   \State unconverged=true;
   \While {unconverged}
       \State $r(i,j) \gets$ equation~(\ref{eq:responsibility});                               \Comment update responsibilities.
       \State $a(i,j) \gets$ equations~(\ref{eq:availab1}) and~(\ref{eq:availab2});          \Comment update availabilities.
       \State $\hat{c}_i \gets$ equation~(\ref{eq:assignments});                            \Comment compute cluster assignments.
       \State unconverged $\gets$ check convergence;
   \EndWhile
   \State $Ind \gets \hat{c}_i$;   \Comment get indices of exemplar priors.
   \State \Return ($Ind$);
\EndFunction
\end{algorithmic}
\end{algorithm}

\subsection{Consolidation and Reconstruction}
\label{sec:Reconstruction}
Once the exemplar priors are obtained from the affinity propagation algorithm, they can be used to consolidate and reconstruct raw point clouds same as in~\cite{Gal2007}, see Algorithm~\ref{Shape Reconstruction alg}.

Given an input raw point scan $\mathcal P$, we construct a set of local neighborhoods $\emph{Nb}$ from $\mathcal P$, as we process the initial priors in Section~\ref{Pre-Processing}.
We select a set of local centers by picking a first point as a seed and search its neighboring point set within a bounding sphere of radius $R$, and we label the neighboring points as covered and continue with the same manner until all points are covered. As a result, the set of local neighborhoods $\emph{Nb}$ is obtained. We transform them to a canonical placement using the weighted PCA positioning explained before and store them for later use.
Subsequently, the shape descriptor $V(\emph{Nb})$ is calculated for each transformed local neighborhood using Equation~\ref{descrip eq}.

Once the shape descriptors of all local neighborhoods within $\mathcal P$ are obtained, we search for each local neighborhood $\emph{Nb}$ the most similar exemplar prior from $\mathcal{L}^{\star}$ by computing the Euclidean distances between their corresponding geometric moments.
To find more reliable matches, we do not just rely on the distance between shape descriptors, but also use a more flexible matching refinement procedure,
which proceeds on the basis of the confidence score metric defined with the mean squared error of the point distances~\cite{Gal2007}.
Thus, the similar priors can be reliably found for all the local neighborhoods within the input scan.

After finding the corresponding exemplar prior for each local neighborhood, we apply the inverse of the transformations stored before to each matched exemplar prior, and then align them into the input data $\mathcal P$.
The Iterative Closest Point (ICP) registration technique~\cite{Rusinkiewicz2001} is then performed to refine the alignment of matched priors to $\mathcal P$.
Consequently, the raw point scan is consolidated and thus the data quality is significantly improved.

Having the consolidated point cloud, we exploit the similar reconstruction method in~\cite{Gal2007} based on the moving least squares technique~\cite{Levin2003}.
As a result, we are able to produce geometrically faithful surfaces, where sharp features and fine details are well preserved.

\begin{algorithm}
\algorithmicrequire Raw Point Scan $\mathcal P$ , Exemplar Priors $\mathcal{L}^{\star}$ .\\
\algorithmicensure Reconstructed Surface $S$.
\caption{Surface Reconstruction} \label{Shape Reconstruction alg}
\begin{algorithmic}
\Function {Surface Reconstruction} {$\mathcal P$,$\mathcal{L}^{\star}$}
    \State $C \gets$ subset$(\mathcal P)$;                         \Comment centers of the local neighborhoods.
    \For{$i = 1$ to size ($C$)}
          \State $c\gets C(i)$;
          \State $Nb(c)\gets$ equation~\ref{Prior eq};                     \Comment generate the local neighborhood.
          \State Weighted PCA ($Nb(c)$);                                   \Comment store the transformation.
          \State Canonical scaling ($Nb(c)$);                              \Comment store the scale.
          \State $V(Nb(c))\gets$ equation~(\ref{descrip eq});              \Comment geometric moment for the local neighborhood.
          \State $Index \gets$ ANN ($V(Nb(c)),V(\mathcal{L}^{\star}$));    \Comment nearest exemplar prior.
          \State $Pr \gets$ Inverse transformation of $\mathcal{L}^{\star}(Index)$;
          \State $Pr \gets$ ICP( $Nb(c)$,$Pr$);
          \State $\mathcal P \gets \mathcal {P}+Pr$;                       \Comment augmented point cloud.
    \EndFor
    \State $S \gets$ MLS$(\mathcal P)$;                                    \Comment moving least square.
    \State \Return ($S$);
\EndFunction
\end{algorithmic}
\end{algorithm}

\begin{figure*}[!t] \centering
  \includegraphics[width=\linewidth]{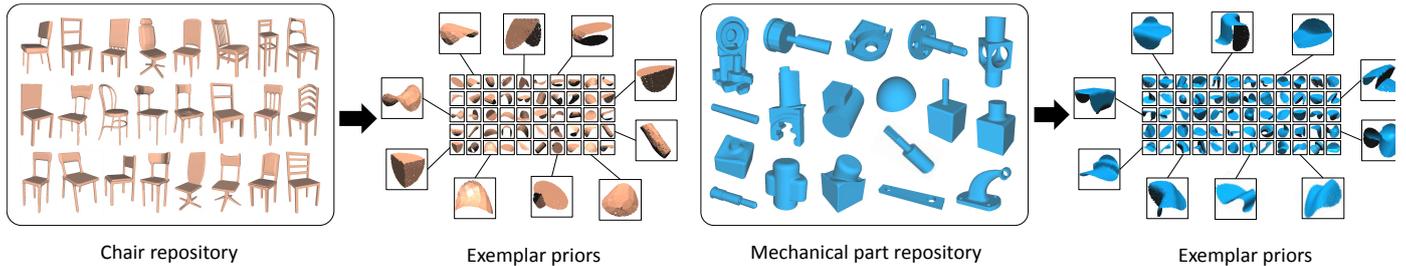}\\
  \caption{A set of exemplar priors for chair and mechanical part datasets. In the chair database, there are a total of $24$ chairs with different design styles. Using our algorithm, $208$ exemplar priors are obtained. Theoretically, those $208$ representative priors can be used to reconstruct any type of chairs with the same style. For the mechanical parts, we choose $18$ model and generate $196$ exemplar priors. For ease of exposition, we illustrate only $50$ exemplar priors for both datasets.}
  \label{fig:Exemplar_Priors}
\end{figure*}
\begin{figure*}[!t]
  \centering
  \includegraphics[width=\linewidth]{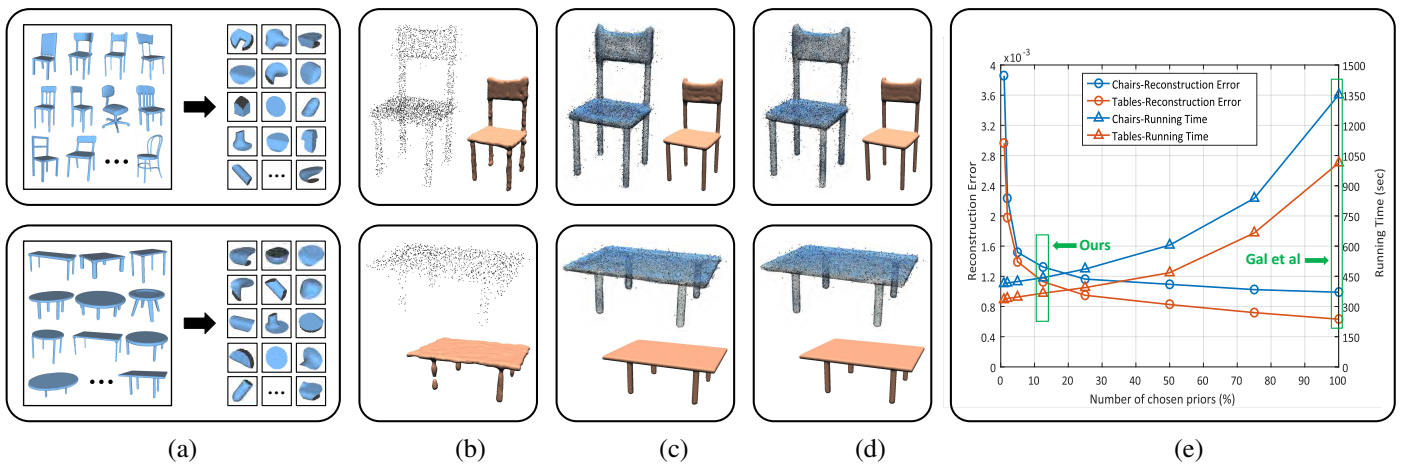}\\
  \hspace{19 mm}(a)\hspace{33 mm} (b) \hspace{21 mm} (c)\hspace{21 mm} (d)\hspace{37 mm} (e)\hspace{23 mm}
  \caption{Comparison to Gal et al.~\cite{Gal2007}.  (a) Training datasets with a set of learned exemplar priors.  (b) Input point data and the reconstruction result from Poisson. (c) Augmentation of the input point data using the \emph{exemplar} priors and the reconstructed model from our method. (d) Augmentation of the input cloud using \emph{all} priors in the library and the reconstructed result from Gal et al. (e) The reconstruction errors and the running time in terms of different numbers of exemplar priors.}
  \label{fig:Priors_Number_k}
\end{figure*}
\section{Results}
\label{sec:results}

In this section, we run our reconstruction method on a variety of raw point clouds, and we validate its effectiveness in terms of feature preservation and detail recovery.
In order to demonstrate the power of our exemplar priors (Figure~\ref{fig:Exemplar_Priors} illustrates the exemplar priors obtained for chair and mechanical part datasets), we comprehensively compare our method to the other related methods on both benchmark data and raw point scans. Additionally, we perform quantitative evaluations to measure the reconstruction quality of our algorithm.

\subsection{Parameter Effects}

\textbf{Number of chosen priors.}
Figure~\ref{fig:Priors_Number_k} shows the effect of the number of chosen priors on the reconstruction quality and the running time.
Given two collections of 3D models, we generate their respective prior libraries.
The input raw scans in Figure~\ref{fig:Priors_Number_k}~(a) are sparse and contain a certain level of noise.
Using our algorithm, we can plot the reconstruction errors and the running time in terms of the different numbers of exemplar priors. Note that our method learns only a small set of exemplar priors (around 5\% of the prior library size), while the reconstruction in Gal et al.~\cite{Gal2007} utilizes all priors from the library.
Based on all original priors, apparently, the reconstruction quality is the best and thus the error is the lowest. However,
The priors generated from a collection of 3D models of the same class could have many duplicate priors that share similar geometric properties, thus the number of priors presented in the library could be huge, which makes the matching process quite time-consuming.
As can be seen in Figure~\ref{fig:Priors_Number_k}~(e), when the number of chosen priors decreases, the reconstruction time decreases and the error becomes higher.
In contrast to~\cite{Gal2007}, our method makes a good trade-off between the number of exemplar priors $k$ and the reconstruction quality, and consolidates input clouds based on a limited number of priors, which offers more efficient priors matching, resulting in faster reconstructions up to three orders of magnitude compared to~\cite{Gal2007}.
When using the set of exemplar priors learned by our algorithm, there is a good balance between the reconstruction quality and the running time.
This suggests that we are able to reconstruct a geometrically faithful models from raw point data with much less difficulty and time, even though the scanned model is geometrically and topologically complex.

\begin{figure}[!t]
  \centering
  \includegraphics[width=\linewidth]{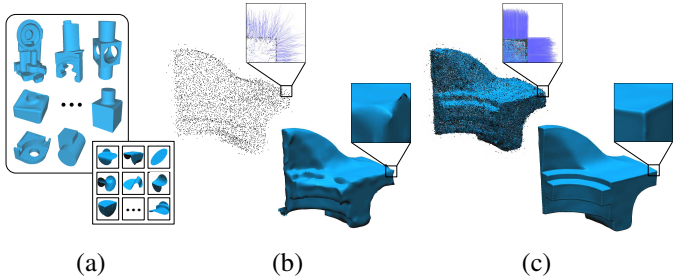}\\
  \hspace{3mm}(a)\hspace{21mm} (b) \hspace{27 mm} (c)\hspace{10 mm}
  \caption{Preserving sharp features on the \emph{Fandisk} model.  (a) Training dataset with a set of learned exemplar priors.  (b) Input point data with $10\%$ Gaussian noise and the reconstruction result using RIMLS. (c) Augmentation of the input point data using the exemplar priors and The reconstructed model from our method. Comparatively, the sharp features are better preserved by our method.}
  \label{fig:Sharp_Fan}
\end{figure}

\textbf{Prior size.}
Our reconstruction method has one major parameter: priors radius $R$, which affects the reconstruction quality and the computation time.
Note that the value of the radius $R$ should be chosen depending on the size of the models present in the dataset.
If the value of $R$ is too big, then the priors possess more meaningful semantics with a small number of redundant elements, resulting in a big number of exemplar priors so that the learning phase is a bit time-consuming. In addition, the local neighborhoods may not find their similar exemplar priors in the matching process, which can definitely harm the reconstruction quality.
On the contrary, if the value of the radius is too small, the number of priors in the library could be huge, while most of them look like a flat disk, leading to a small number of exemplar priors. As a result, both the learning and the reconstruction phases are quite time-consuming.
We choose $R$ between 0.05 and 0.1 times the diagonal of the bounding box of each 3D model.
In practice, We have conducted a number of testings on different input scans, and found that the values of $R$ in the interval [0.05, 0.1] times the diagonal of the bounding box of each 3D model always produce pleasing results in all experiments.
For datasets that are very similar to the shape being reconstructed (e.g., humans and armadillos), we set $R=0.1$. For the datasets that undergo some geometric variations, we fix $R$ at $0.05$.

\begin{figure}[!t]
  \centering
  \includegraphics[width=\linewidth]{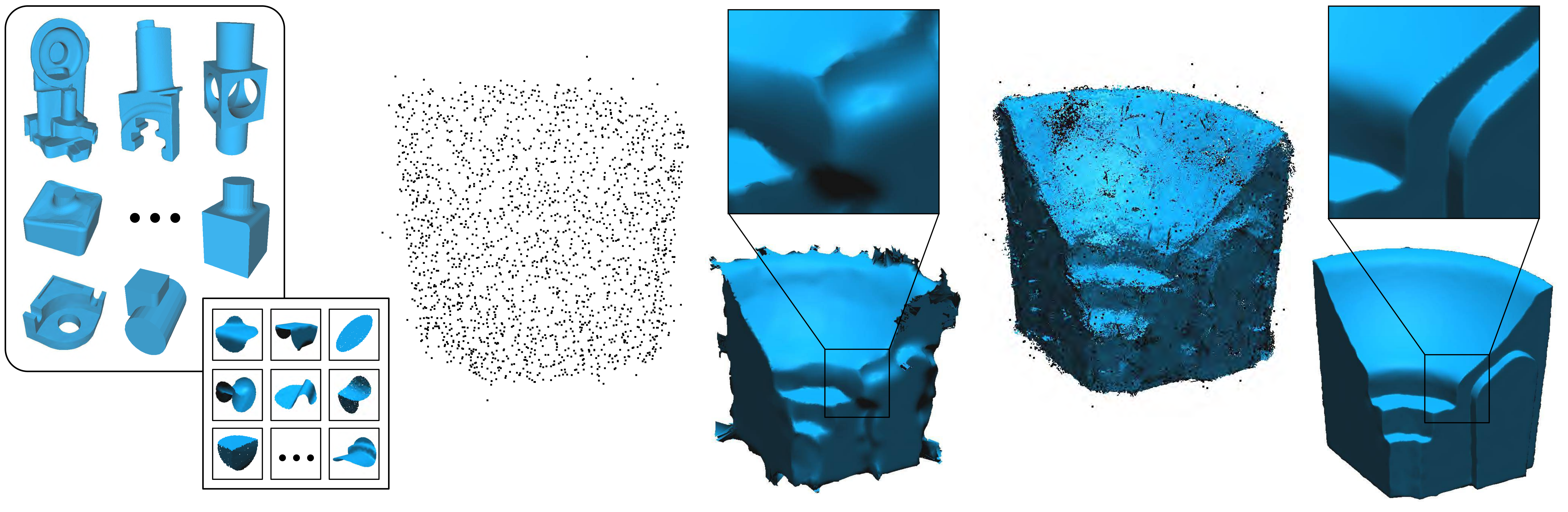}\\
  \hspace{6mm}(a)\hspace{24mm} (b) \hspace{27 mm} (c)\hspace{10 mm}
  \caption{Preserving sharp features on the model of a \emph{Mechanical part}. (a) The training mechanical part dataset and a bench of exemplar priors learned by our algorithm. (b) The raw scanned point data with $10\%$ Gaussian noise and the reconstruction result from RIMLS. (c) The augmented point data and the reconstruction result from our method. The close-up views clearly show that our method is capable of recovering sharp features even in the presence of noise.}
  \label{fig:Sharp_Mech_Part}
\end{figure}

\begin{figure*}[!t] \centering
  \includegraphics[width=\linewidth]{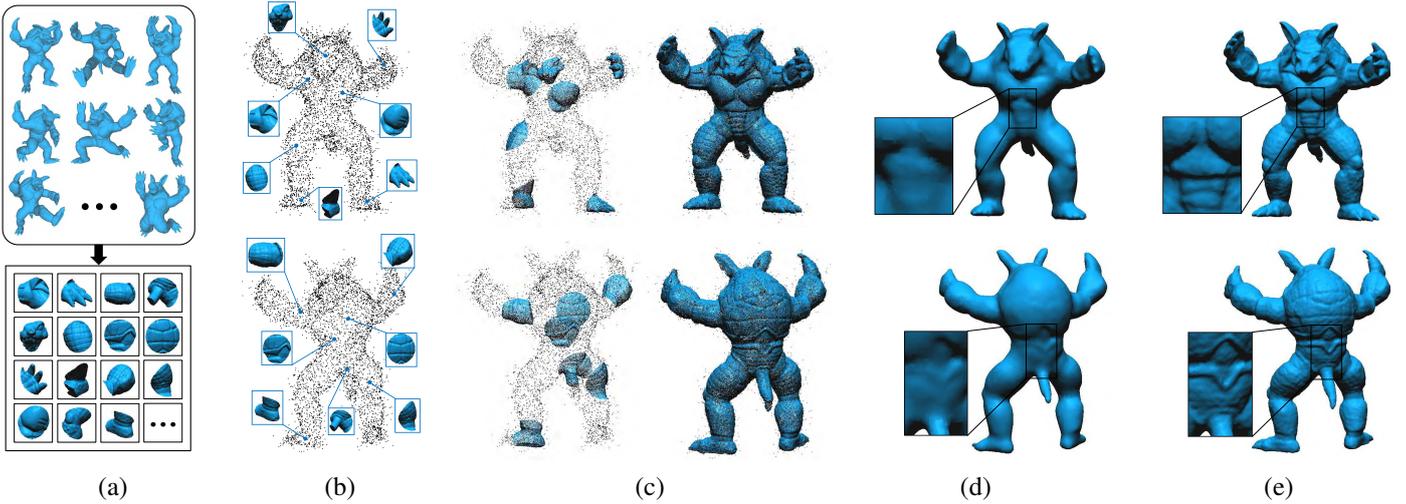}\\
  \hspace{0 mm}(a)\hspace{25 mm} (b) \hspace{35 mm} (c)\hspace{38 mm} (d)\hspace{35 mm} (e)\hspace{0 mm}
  \caption{Detail recovery on the \emph{Armadillo} model. The first and second rows represent the front and the back views respectively. (a) The armadillo database used to learn the exemplar priors. (b) The input noisy raw data with a set of learned priors matched with their local neighborhoods. (c) The alignment of the exemplar priors into the input point cloud, where initial alignments are represented on the left and the final augmented point cloud is shown on the right. (d) Poisson reconstruction result of the input point cloud. (e) Our reconstruction result. Poisson reconstruction lacks significant details while our method successfully recovers fine details.}
  \label{fig:Details_Armadillo}
\end{figure*}

\subsection{Reconstruction}

\textbf{Sharp feature preservation.}
Figure~\ref{fig:Sharp_Fan} compares the reconstruction results on the fandisk scan from RIMLS~\cite{Oztireli2009} and our method.
As can be seen in Figure~\ref{fig:Sharp_Fan}(a), our algorithm is able to find the exemplar priors that can represent the mechanical part dataset. Using these priors to match and augment the local neighborhoods of the input scan can inherently improve the quality of the input cloud and preserve the sharp features in Figure~\ref{fig:Sharp_Fan}(c), while the sharp features are highly damaged by RIMLS method in Figure~\ref{fig:Sharp_Fan}(b).
From the close-up views, the normals of the points around the sharp edges are well recovered with our approach, since they are originated from the triangular meshes.
Figure~\ref{fig:Sharp_Mech_Part} shows the reconstruction result of a mechanical part using our method.
We run RIMLS on the input scan data to get the reconstruction result in Figure~\ref{fig:Sharp_Mech_Part}(b), from which sharp features are severely blurred. By learning from a set of mechanical models, our algorithm chooses a set of exemplar priors in Figure~\ref{fig:Sharp_Mech_Part}(a), based on which the low-quality scan is pleasingly consolidated. The reconstruction result of the enhanced input cloud is produced in Figure~\ref{fig:Sharp_Mech_Part}(c), where all sharp edges are well preserved.

\textbf{Detail recovery.}
To demonstrate the effectiveness of our method in terms of detail recovery, we generate the surface models from the input scans using our algorithm and the Poisson reconstruction method on the armadillo and the chair models.
Figure~\ref{fig:Details_Armadillo} demonstrates how our method can effectively recover small details from noisy raw point data.
Given a dataset of armadillo model from~\cite{Chen2009,Jacobson2012}, we run our algorithm to learn a set of exemplar priors ( Figure~\ref{fig:Details_Armadillo}(a) illustrates only 15 priors from the 324 learned priors), priors matching and point cloud augmentation procedures are shown in Figures ~\ref{fig:Details_Armadillo}(b) and ~\ref{fig:Details_Armadillo}(c) respectively.
The reconstructed surface using Poisson method in Figure~\ref{fig:Details_Armadillo}(d) from the input scan lacks significant details as shown in the zoom-in views.
Comparatively, our method succeeds in recovering fine features based on the exemplar priors.

\begin{figure}[!t] \centering
  \includegraphics[width=\linewidth]{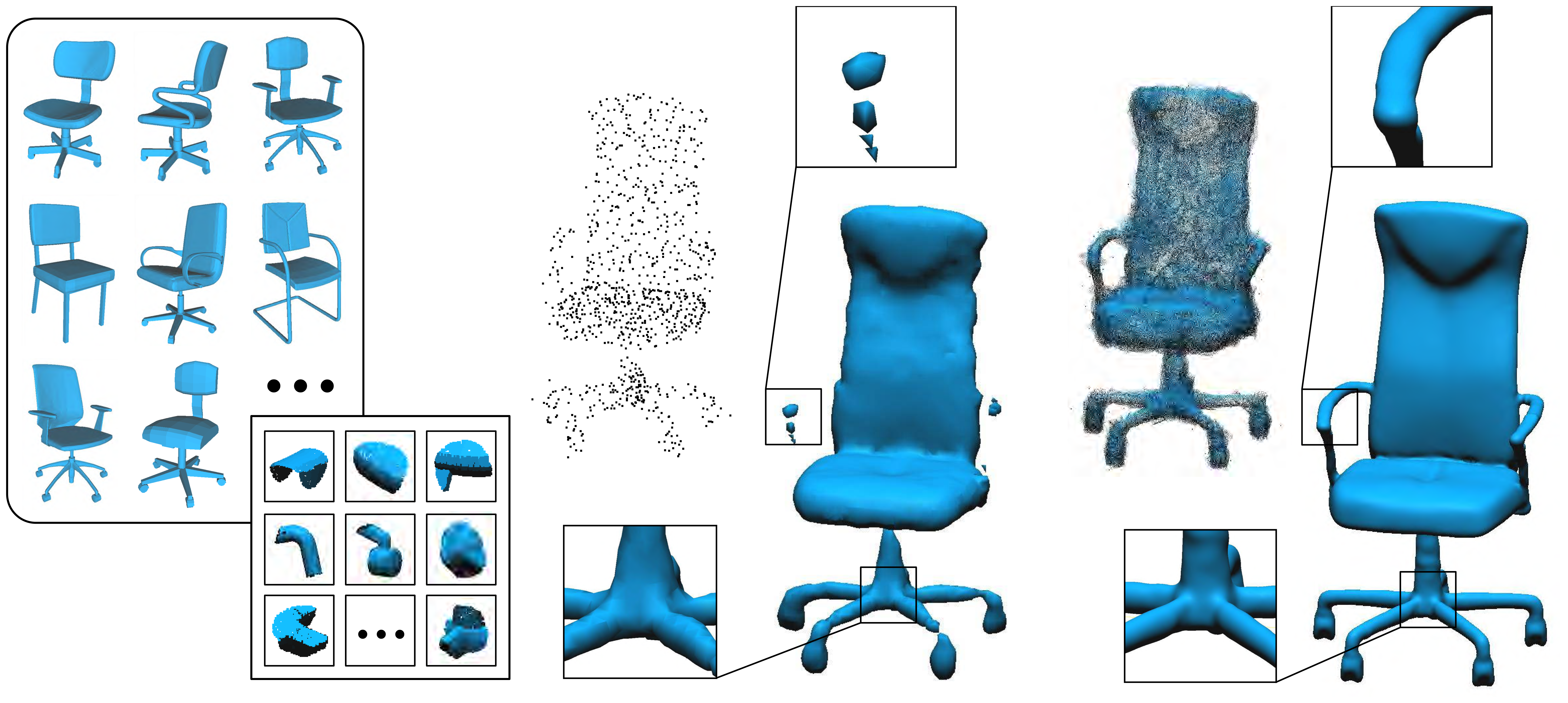}
  \hspace{20 mm}(a)\hspace{25 mm} (b) \hspace{24 mm} (c)\hspace{0 mm}
  \caption{Detail recovery on the \emph{Chair} model. (a) The chair database used to learn the exemplar priors. (b) The Poisson reconstruction result on the input raw points. (c) The reconstruction result using our method. Compared to the poisson reconstruction, our method can recover more details benefiting from the fine details carried by the exemplar priors, as demonstrated in zoom-in views.}
  \label{fig:Details_Chair}
\end{figure}

\begin{figure}[!t] \centering
  \includegraphics[width=\linewidth]{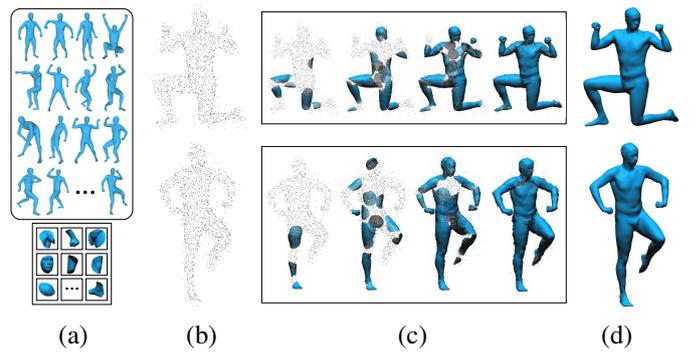}
  \hspace{10 mm}(a)\hspace{12 mm} (b) \hspace{22 mm} (c)\hspace{22 mm} (d)\hspace{0 mm}
  \caption{Dealing with distinct poses. (a) The training human database with various poses and a set of learned exemplar priors. (b) The input point scans. (c) The augmentation procedure on the input point data. (d) Our reconstruction results. Although the corresponding poses of the input scans are not included in the database, our exemplar priors still can effectively consolidate them and reconstruct geometrically faithful models.}
  \label{fig:Human_pose}
\end{figure}

Figure~\ref{fig:Details_Chair} illustrates the reconstruction results on a chair model.
We run the poisson reconstruction method on the input scan, where the surface is bumpy and important details are missing,
as shown in Figure~\ref{fig:Details_Chair}(b).
By learning from a dataset of chair models ~\cite{Alhashim2015,Huang2015}, our algorithm chooses a total of 116 exemplar priors (8 of these priors are demonstrated in Figure~\ref{fig:Details_Chair}(a)).
Using those exemplar priors, our method is capable of recovering fine details, even from a sparse and noisy raw scan,
as demonstrated in Figure~\ref{fig:Details_Chair}(c).

\textbf{Pose invariant.}
Another important characteristic of our learned exemplar priors is that they can deal with models from the same class of objects with different poses, in another word, they can be used to reconstruct a new model with a new pose different from those used in the learning process.
Figure~\ref{fig:Human_pose} shows the reconstruction results on human models with distinct poses from SCAPE database~\cite{Scape2005}.
We train our algorithm on 30 human models of the same person with a variety of poses, as a result, $296$ exemplar priors are obtained to be able to represent the human model, as can be seen in Figure~\ref{fig:Human_pose}(a).
Given sparse and noisy input scans of human models in Figure~\ref{fig:Human_pose}(b), we use the learned priors to consolidate and augment the input clouds.
Figure~\ref{fig:Human_pose}(c) illustrates $4$ stages of the augmentation procedure where the exemplar priors are matched and aligned to their corresponding local neighborhoods.
The generated dense point clouds contain faithful normals since they are originated from the input mesh models, and therefore we are able to reconstruct geometrically faithful surface models, where the important details can be well recovered, see Figure~\ref{fig:Human_pose}(d).

\begin{figure*}[!t] \centering
  \includegraphics[width=0.85\linewidth]{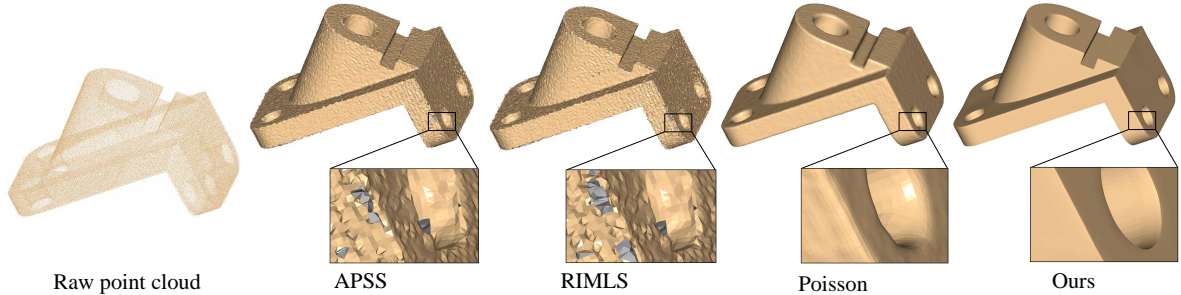}
  \caption{Reconstruction comparison on the \emph{Anchor} model from APSS(resolution=$320^3$), RIMLS (resolution=$320^3$), Poisson (depth=12) and our method. APSS and RIMLS can preserve sharp features to some extent, while a number of artifacts are yielded in the results. Poisson produces the favorably smooth model, where sharp features are blurred. In contrast, our result is relatively better.}
  \label{fig:Anchor_45}
\end{figure*}

\begin{figure*}[!t] \centering
  \includegraphics[width=0.85\linewidth]{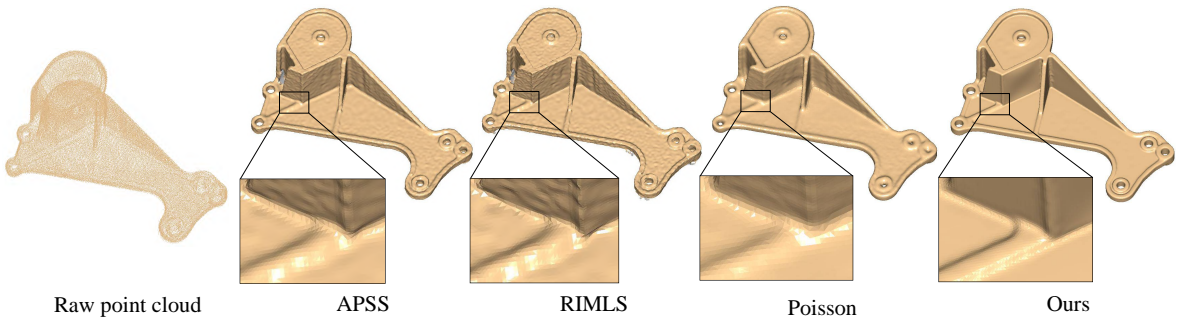}
  \caption{Reconstruction comparison on the \emph{Daratech} model from APSS (resolution=$320^3$), RIMLS (resolution=$320^3$), Poisson (depth=12) and ours . Our method outperforms the others in terms of sharp feature preservation in the face of noise.}
  \label{fig:Daratech_37}
\end{figure*}

\subsection{Comparisons to state-of-the-art reconstruction techniques}
To demonstrate the effectiveness of our method,
we compare it to the state-of-the-art reconstruction techniques, including RIMLS~\cite{Oztireli2009}, APSS~\cite{Guennebaud2007} and Poisson~\cite{Kashdan2006}.
We try our best to fine-tune the parameters for all those methods in order to achieve the best experimental results.
The experimental data include surface reconstruction benchmark data~\cite{Berger2013} (Anchor and Daratech point clouds),
as well as the raw scanned data of two mechanical parts taken from~\cite{GlobFit2011}.

\textbf{Benchmark data.}
Figure~\ref{fig:Anchor_45} compares the reconstruction results of the Anchor model with \emph{sharp features}.
The input point data contain a certain level of noise.
RIMLS and APSS are able to retain sharp features to some extent, while there are some bumpy artifacts.
Moreover, a number of small holes exist in their results due to noise.
Poisson yields a smooth, closed surface thanks to its reconstruction characteristics.
However, sharp features are severely blurred.
In contrast, our method consolidates the input point data based on the exemplar priors learned from a set of mechanical part models,
and hence produces a quality surface, where sharp features are well preserved.
Figure~\ref{fig:Daratech_37} presents the comparison results of the Daratech model with \emph{small details}.
APSS, RIMLS and our method obtain relatively better results than Poisson does in terms of retaining sharp features.
Note that notable defects (e.g. holes) are still generated from APSS and RIMLS.
As highlighted, RIMLS and ours preserve more details than APSS and Poisson do.
However, RIMLS ruins the small features.
Comparatively, only ours achieves more favorable results,
in which fine details and sharp features are properly recovered.

\begin{figure*}[!t] \centering
  \includegraphics[width=0.85\linewidth]{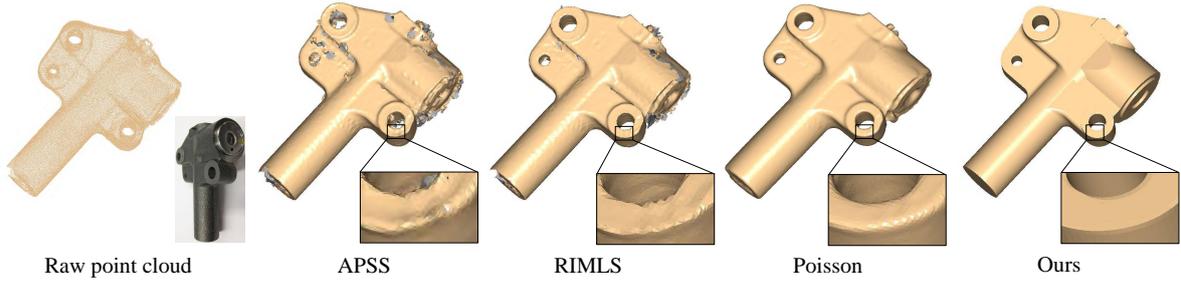}
  \caption{Reconstruction comparison on the raw scan of the \emph{Clutch-part} model from APSS (resolution=$350^3$), RIMLS (resolution=$350^3$), Poisson (depth=12) and ours.}
  \label{fig:clutch_part}
\end{figure*}

\begin{figure*}[!t] \centering
  \includegraphics[width=0.85\linewidth]{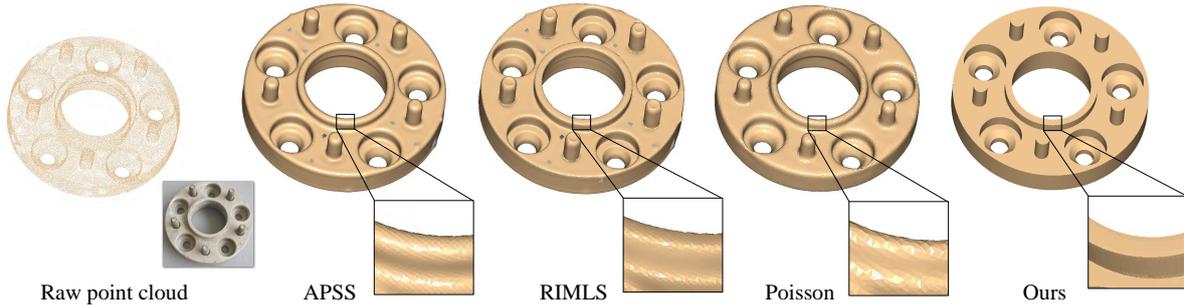}
  \caption{Reconstruction comparison on the raw scan of the \emph{Wheel} model from APSS (resolution=$320^3$), RIMLS (resolution=$320^3$), Poisson (depth=12) and ours.}
  \label{fig:wheel_model}
\end{figure*}

\textbf{Raw scans.}
Figure~\ref{fig:clutch_part} shows the comparison results of the Clutch part with \emph{raw point data}.
The raw point cloud contains some noise, and the points are distributed non-uniformly.
Moreover, the data on some regions are fairly sparse.
Due to noise, APSS and RIMLS generate a number of artifacts (e.g. burrs),
while Poisson obtains a visually pleasing surface.
As shown in the close-up views, the edges of the cylindrical hole are collapsed by APSS and RIMLS.
Poisson smooths out those sharp edges, which, however, are well preserved by our method.
Figure~\ref{fig:wheel_model} compares the reconstruction results of the Wheel model with \emph{raw point data}.
In contrast, our method outperforms the other three methods in terms of recovering sharp features,
even for a fairly noisy scan.

\subsection{Quantitative Evaluations}
The results shown above have visually demonstrated the superiority of our method regarding sharp feature preservation and detail recovery on raw scanned points.
In this section, we perform quantitative evaluations on our reconstruction approach.
We first show the Hausdorff distance histograms of the reconstruction results from Figures~\ref{fig:Anchor_45}-\ref{fig:wheel_model}.
We also investigate the robustness of our method to different levels of noise and sparsity in a quantitative manner.
Raw scan data always suffer from varying sampling resolution, noise, and sparsity.
While it is difficult to exactly simulate real scan data,
we incorporate the synthetic experiments to validate the robustness.

\begin{figure*}[!t] \centering
  \includegraphics[width=1\linewidth]{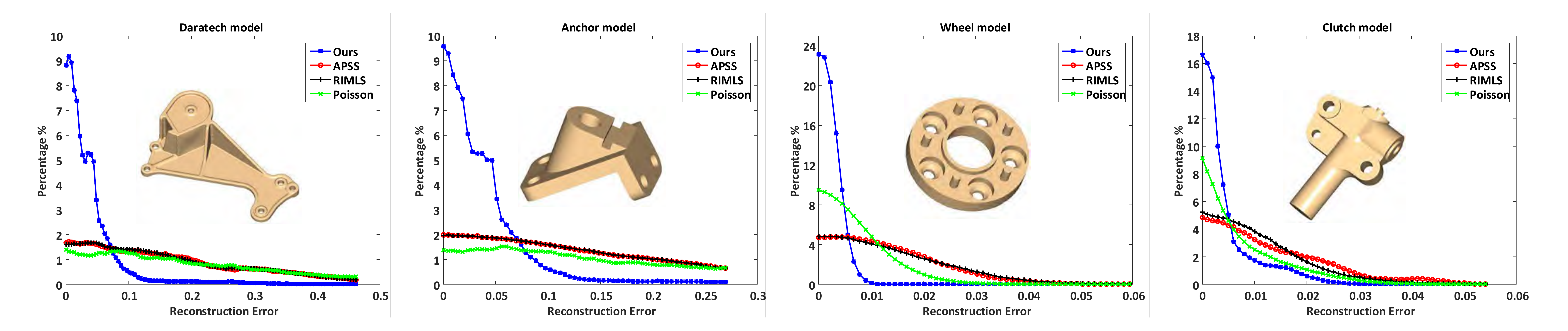}
  \caption{The histograms show the Hausdorff distances between the reconstructed models from the compared methods and the original models. The horizontal axis is the reconstruction error between the reconstructed model and the original one, and the vertical axis stands for the corresponding percentage to each error value.}
  \label{fig:Distances_with_SRB}
\end{figure*}

\textbf{Hausdorff distance.}
To demonstrate the geometric fidelity of the reconstructed surface relative to the original scan,
the Hausdorff distance between the original scan and the surface mesh is computed.
Figure~\ref{fig:Distances_with_SRB} shows the detailed comparisons of the Hausdorff distance results of Figures~\ref{fig:Anchor_45}-\ref{fig:wheel_model}.
The horizontal axis is the absolute distance value between the reconstructed mesh and the original scan,
and the vertical axis is the corresponding histogram (in percentage) with respect to each distance value.
From the histograms, we can see that our method yields shorter Hausdorff distances,
indicating that our method produces more geometrically closer surfaces to the original models.

\begin{figure*}[!t] \centering
  \includegraphics[width=0.7\linewidth]{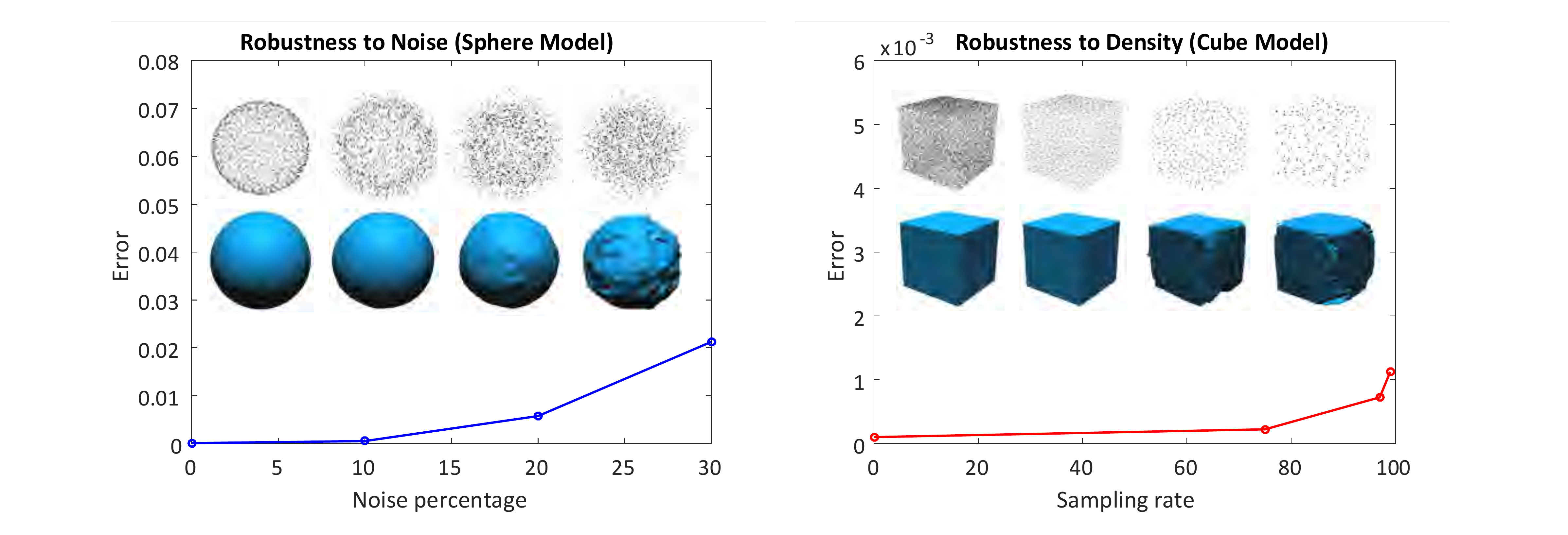}
  \caption{Reconstruction robustness in terms of noise and density. Here we choose the appropriate priors to reconstruct the sphere point clouds with increasing level of noise (left) and the cube scans with different sampling rates (right). From those tendencies, we note that our reconstruction approach is quite robust to reasonably high levels of noise and sparsity.}
  \label{fig:Noise_Density}
\end{figure*}

\textbf{Robustness to noise and sampling rates.}
To further evaluate the robustness of our reconstruction algorithm to noise and sampling rates, we choose two mesh models (the sphere and the cube) as ground truth, we perform virtual scanning to generate the point data, we reconstruct these point data by using the adequate priors and thus correctly measure the precision to the ground truth, as demonstrated in Figure~\ref{fig:Noise_Density}.
For each scan, we successfully reconstruct the surface and compute the reconstruction error between the surface and the ground truth using Equation~\ref{Error eq}. In such a way, we are able to plot the reconstruction precision of the two models with respect to the level of noise (left) and the sampling rates (right).
As observed, with a reasonably high level of noise and outliers, our prior-based reconstruction technique still achieves a high precision. As the artifacts intensify and the corruption becomes extremely severe, the performance of our method could drop considerably.
By increasing the sampling ratio of the original points, our approach still obtains a good reconstruction precision. As the level of sampling becomes fairly high, our matching-to-alignment procedure may fail to find the appropriate priors for the local neighborhoods, which may cause wrong matching as shown in the third and fourth columns.

\subsection{Performance}
We have implemented the technique presented in the previous sections in C++ and all experiments are performed on a PC with a 2.4GHz CPU and 6.0 GB of RAM.
The total running time of our algorithm can be divided into two parts: the exemplar priors learning and the reconstruction phases.
Priors learning phase includes the construction of the prior library and the affinity propagation clustering method, this phase is quite fast and it depends on the number of the input shapes and the number of priors in the library ( the learning time increases as the number of prior library increases). The second phase of our algorithm is relatively slow and it relies on the input point scan (please note that the reconstruction phase includes the local neighborhoods construction, matching and alignment procedures for each local and the MLS surface reconstruction).
Overall, the time performance of our method is still comparatively efficient.
Table~\ref{tab:timings} gives the running time performance of our method and some of the crucial parameters used in our experiments.

\subsection{Limitations}
Our method is expected to behave well with different shape categories, meanwhile there are a few limitations that have to be discussed so far.
Our algorithm fails when dealing with more challenging repositories with small number of redundant elements, such as complex organic shapes.
In addition, if there are large holes within the input scans or big missing parts, our method may fail to complete them based on the ``matching-to-alignment'' strategy.

\section{Conclusion and Future Work}

In this paper, we present a pipeline for surface reconstruction on the basis of learning local shape priors from a database of 3D shapes.
Our algorithm exploits the self-similarity patches extracted from a database of 3D shapes within the same class, uses them to enhance the imperfect input scans from the same class, and reconstructs a faithful surface.
Our reconstruction framework can be generalized to handle various categories of 3D objects which may have different geometrical and topological structures.
Instead of tuning magic threshold parameters of the surface reconstruction techniques, our approach gives the user an intuitive way to extract the priors from existing models for surface reconstruction, which does not depend on any pre-defined parametrical model or heuristic regularities.
A variety of experimental results on synthetic and raw point scans demonstrate that our method is capable of producing quality surfaces,
where sharp features and fine details are recovered well, in the presence of reasonably high level of noise, outliers and sparsity.
The comprehensive experiments also demonstrate that the set of \emph{exemplar priors} learned from a shape database is a powerful tool to elegantly produce consolidated scans from poor and noisy point clouds, that can be used for several applications.

In the future, we plan to focus on developing the algorithm to handle a variety of problems such as filling holes and dealing with more challenging input point scans.
Another interesting research direction is to mine the global, where high-level information can be extracted from the training database models such as local prior graphs, and then integrated with the local information of the priors to handle various geometry processing applications.

\begin{table}[!t]
\def\arraystretch{1.2}
\centering
\caption{Running time statistics and the parameters used in our experiments. Pts: Number of points of the input scan, M: Size of shape repository, $k$: Number of exemplar priors, $R$: Relative radius value, L Time: Priors learning time and R Time: Reconstruction time. }
\label{tab:timings}
\begin{tabular}{lcccccc}
\multicolumn{1}{l}{\textbf{Figure}} & \multicolumn{1}{c}{\textbf{Pts}} & \multicolumn{1}{c}{\textbf{M}} & \multicolumn{1}{c}{\textbf{$k$}} & \multicolumn{1}{c}{\textbf{$R$}} & \multicolumn{1}{c}{\textbf{L Time}} & \multicolumn{1}{c}{\textbf{R Time}} \\ \hline
Fig~\ref{fig:Sharp_Fan}           & 3999              & 36             & 346            & 0.05            & 4m01s           & 6m47s \\ \hline
Fig~\ref{fig:Sharp_Mech_Part}     & 2999              & 36             & 346            & 0.05            & 4m01s              & 6m11s    \\ \hline
Fig~\ref{fig:Details_Armadillo}   & 13999             & 10             & 324            & 0.1             & 4m54s              & 13m02s   \\ \hline
Fig~\ref{fig:Details_Chair}       & 3499              & 22             & 116            & 0.05            & 2m17s              & 5m23s    \\ \hline
Fig~\ref{fig:Human_pose}~(1)       & 5999              & 30             & 296            & 0.1             & 3m10s              & 7m53s    \\ \hline
Fig~\ref{fig:Human_pose}~(2)       & 5999              & 30             & 296            & 0.1             & 3m10s              & 7m39s    \\ \hline
Fig~\ref{fig:Anchor_45}           & 107818            & 36             & 346            & 0.05            & 4m01s              & 12m25s   \\ \hline
Fig~\ref{fig:Daratech_37}         & 84682             & 36             & 346            & 0.05            & 4m01s              & 12m08s   \\ \hline
Fig~\ref{fig:clutch_part}         & 84519             & 36             & 346            & 0.05            & 4m01s              & 11m41s   \\ \hline
Fig~\ref{fig:wheel_model}         & 35565             & 36             & 346            & 0.05            & 4m01s              & 9m06s    \\ \hline
\end{tabular}
\end{table}

\section*{Acknowledgements}
{We thank the anonymous reviewers for their valuable comments and suggestions.
The work was supported in part by National Natural Science Foundation of China (61402224),
the Fundamental Research Funds for the Central Universities (NE2014402, NE2016004),
the Natural Science Foundation of Jiangsu Province (BK2014833),
the NUAA Fundamental Research Funds (NS2015053),
and Jiangsu Specially-Appointed Professorship.
}

%

%






%
%
%
%
\end{document}